\documentclass{ws-procs9x6}

\begin{document}

\title{QCD Effective Actions at High Temperature and the Quasiparticle
Picture}

\author{Cristina Manuel \footnote{\uppercase{W}ork 
supported by  the \uppercase{G}eneralitat
\uppercase{V}alenciana, grant CTIDIA/2002/5.}}

\address{Instituto de F\'{\i}sica Corpuscular\\
Universitat de Val\`encia -C.S.I.C. \\
Edificio Institutos de Paterna,\\ Apt. 2085, 46071 Val\`encia. Spain \\
E-mail: Cristina.Manuel@ific.uv.es}


\maketitle

\abstracts{Transport theory is an efficient approach to derive an
effective theory for the soft modes of QCD at high temperature.
It is known that the leading order operators of this theory
can be obtained from (semi-classical) kinetic equations of quasiparticles
carrying  classical or quantum color charges.
Higher order operators can also be
obtained. Discrepancy between these quasiparticle models starts
for dimension 4 operators, which  converge in the limit of
high dimensional color representations. These  models
are reviewed and compared.}

\section{Introduction}

The study of the high temperature phase of QCD is necessary to characterize
the properties of the quark-gluon plasma (QGP), 
relevant for the phenomenology of
 heavy ion collisions.
A serious obstacle to make robust predictions of its properties
is the non-perturbative character of the long distance physics. 
The standard Monte Carlo techniques can only be used to
get the static properties of the system, either by performing 4d lattice
or (dimensional reduced)  3d lattice simulations.
Unfortunately, a  similar powerful approach to study the dynamics
in  this phase is still missing. Efforts to develop  non-perturbative 
tools for the QGP  are certainly required.

One approach to study dynamical properties of non-Abelian plasmas is based on
the fact that the low energy modes behave classically. It is possible to
construct effective classical field theories which are amenable to numerical
treatment. These theories treat the hard modes as quasiparticles moving in
the background of  soft classical fields, and  they are modeled with a
simple transport equation. Previous sessions of SEWM were intensively
devoted to the 
construction of this program, and I refer to their proceedings for the 
basic references. These effective models allowed one to evaluate
transport coefficients to leading order in $g$, the gauge
coupling constant. 

It seems reasonable to ask at which point the predictions
of kinetic theory depart  from those of quantum field theory. If not, 
it  would be  interesting to see whether the analysis explained above
could be pushed beyond leading order in  $g$.
With this aim I will review the predictions of kinetic theory for systems
close to thermal equilibrium, comparing them with those from quantum field
theory. This approach is constructive, as the intention is not 
to derive transport equations from quantum field theory
(while this is possible!), but rather 
to see if a simple quasiparticle model describes correctly the QCD predictions
for the long distance physics.
This is based on work done in collaboration with Mikko Laine\cite{Laine:2001my}
and Stanis\l aw Mr\' owczy\' nski\cite{Manuel:2002pb}. A similar effort in that direction was first carried
out in Ref.\cite{Bod01}.

There are two different kinetic equations to describe classical
colored particles\cite{Hei83}.
One of them describes color as a classical degree of freedom.
The other one describes color as a quantum degree of freedom,
represented by a matrix  in
a certain representation $R$ of the gauge group $SU(N_c)$. While the two
different kinetic equations predict the same low order operators in these
 effective  theories,  they do not for higher order operators.

\section{Transport Equations for Classical Color}

Let us a consider a  non-relativistic particle of mass $m$ carrying a 
classical color charge $Q^a$, where $a=1$ to $N_c-1$ for a $SU(N_c)$ group.
The Hamiltonian for this particle is
\begin{equation}
 H =
 \frac{1}{2 m} \left( {\bf k} - g {\bf A}^a Q_a \right)^2 + g A_0^a Q_a \ ,
\label{eq:1}
\end{equation}
where $A_\mu^a$ is the non-Abelian vector gauge field.  The color charge is a
dynamical variable, which evolves in time $t$,
 as the canonical momentum ${\bf k}$
and position ${\bf x}$ of the particle. Its phase-space is then 
enlarged by adding the color charge variables.
The Poisson brackets (PB) of this classical system are
also modified\cite{Kel94}.
 Hamiltonian equations of motion can be derived with the
help of these Poisson brackets, and they correspond to the well-known
Wong equations\cite{Wong}.

A statistical description of a system of classical colored particles 
starts  with the definition of the one-particle
distribution function $f({\bf x},{\bf k}, Q;t)$. In the absence of
collisions it evolves according to the Hamiltonian dynamics,
$\frac{d f}{d t} = \left\{ H,  f  \right\}_{PB}$.
A formulation of the transport equation in terms of the canonical momentum
(the phase-space variable) yields a gauge dependent
equation\footnote{The same problem reappears in quantum
field theory derivations of transport equations!}. This problem can be
overcome by rewriting the equation in terms of the (gauge invariant)
kinetic momentum ${\bf p} = {\bf k} -  g {\bf A}^a Q_a$, as this is the
variable associated to the particle velocity  ${\bf v}= {\bf p}/m$. Then
the equation reads
\begin{equation}
v^\mu \left(D_\mu - gQ_a F_{\mu \nu}^a  \frac{\partial}{\partial  p^\nu}
\right)f(x,p,Q)=0\ ,
\label{eq:2}
\end{equation}
where $v^\mu = (1, {\bf v})$, 
 $D_\mu = \partial_\mu - g f^{abc}A_\mu^b Q^c \partial/\partial Q^a$,
and $p_0 = {\bf p}^2/2m$ for a non-relativistic particle.

A Hamiltonian formulation of the dynamics  is very
useful. In particular, it allows one to recognize  constants
of motion,  and thus getting exact solutions
of the collisionless transport equation. If the system is invariant
under space translations in the direction ${\bf n}$, 
then it is easy to prove that ${\bf n}\cdot {\bf k}$ is a constant of motion.
While this is not a gauge
invariant quantity, it is in the type of gauges that are respectful with
the translation invariance of the system,  $n^i \partial_i A_\mu ^a = 0$.
Thus, an exact solution to (\ref{eq:2}) is written as
 $f({\bf n}\cdot( {\bf p} + g {\bf A}_a Q^a))$,
for an arbitrary function $f$, which should be fixed with a boundary
condition.

A generalization of the approach for relativistic systems is straightforward.
All what changes in the final equations is the relativistic, rather than
non-relativistic, dispersion relation. Using this  philosophy 
exact solutions to the relativistic transport equations in static systems
were found in Ref.\cite{Laine:2001my}. The imposed boundary condition
was that $f$ should reduce to the Fermi-Dirac or Bose-Einstein equilibrium
functions when $g \rightarrow 0$, which then fixes the form of $f$. One can
also prove that these distribution functions describe  equilibrium solutions
in the presence of background fields (that is, with no entropy production).
The colored current in the plasma is written in terms of the external fields,
and one can then compute  the associated 
effective action. When compared to the effective action of a system in the
background of static fields as computed in quantum field theory, 
one notices that it matches for the low dimensional operators 
(those proportional to the quadratic and cubic Casimirs),
but there is  a numerical discrepancy in
the coefficient of the dimension 4 operator, unless the particles are in a
high dimensional color representation. It is possible to identify the reason
of this discrepancy in the classical color algebra, and the fact that
the $Q$ charges are commuting objects. Since quarks and gluons are in
low dimensional representations, a quantum treatment of color then
seems mandatory for QCD, at least to reach agreement between the two
theories for all the static operators.

\section{Transport Equations for Quantum Color}

A (first) quantized  treatment of the system described in (\ref{eq:1})
amounts to replace c-numbers by operators in a  Hilbert space.
For a particle with color charge in a representation $R$ of $SU(N_c)$, 
the color charge $Q_a$ is replaced by $T_a$, where $T_a$ is a matrix,
a generator of $SU(N_c)$ in the representation $R$. Poisson brackets are
replaced by commutators. In a Heisenberg representation, one can define
the operators of position, momentum and color charge, and derive their
equations of motion, according to the Hamiltonian dynamics.
 These equations reduce
to  the classical Wong equations  in the limit where all operators commute.

The statistical treatment of the system starts with 
the Wigner operator, defined as a certain Fourier transform of the density
matrix. The Wigner operator is also a matrix in color space, whose dimension
depends on the representation $R$. In a collisionless situation,
 its dynamical evolution is governed by the Hamiltonian. In the limit
where spatial derivatives of higher order can be neglected (that is, performing
a gradient expansion), it reduces to 
\begin{equation}
\left [v^{\mu} D_{\mu}, W( p,x) \right] - {g \over 2}\: v^{\mu}
\left\{ F_{\mu \nu}(x),
{\partial W( p,x) \over \partial p_{\nu}}\right\} =   0  \ ,
\label{eq:3}
\end{equation}
where $D_\mu = \partial_\mu - i g A_\mu^a T^a$.  The equation is the same for
non-relativistic and relativistic systems, all what changes is the
dispersion relation. These transport equations were first derived from
QCD in a gradient expansion in  Refs. \cite{Elz86,Elz87}, while here they have
been obtained from a simple quasiparticle model.

In Ref. \cite{Manuel:2002pb} exact solutions to (\ref{eq:3}) were found for
systems with a translation invariance in the direction $n^\mu$. One
could expect that the form of the solutions to be
$W(p,x) = f(n \cdot (p + g A^a T^a))$. However, this is only the case when
$[ D(n\cdot A), n \cdot A] = 0$. If not, gradient (and non-local) terms
correct this expression. For static systems,
 this implies that an equilibrium solution in
the presence of background fields is only possible if  those are in the
direction of the
Cartan subalgebra of $SU(N_c)$.

One can compare the results arising from a quasiparticle model,
made up of quarks, antiquarks and gluons, and those of QCD.
One finds a perfect matching  in the static limit  with constant
background fields\cite{Gross:1980br}. A discrepancy is  found
when compared with dimensional reduced effective theories\cite{Har00},
if the vector gauge
fields do not commute. There is also a subtlety concerning the gluons.
The transport equation for the hard gluons needs an infrared cutoff, so as
to avoid a double counting of the soft modes (as quasiparticles,
as soft classical fields). This cutoff is also needed in order to avoid
that the distribution function of the hard gluons becomes negative, loosing its
probabilistic interpretation. A matching procedure with the soft classical
field modes allows one to eliminate the cutoff dependence in all final
physical quantities.

\section{Conclusions}

We have pointed out where the classical and quantum color quasiparticle
models agree and disagree. While we have not discussed the structure of the
collision terms, a similar discrepancy beyond leading order in $g$
might be expected there.

I would like to emphasize that  it is worthwhile to explore
these quasiparticle models, not only for the reasons mentioned
in the introduction, but also because 
they might teach us a lot about fluctuations
around global and local equilibrium in a quark-gluon plasma.

\end{document}